\documentclass[sigconf,language=portuguese]{acmart}

\AtBeginDocument{%
  \providecommand\BibTeX{{%
    \normalfont B\kern-0.5em{\scshape i\kern-0.25em b}\kern-0.8em\TeX}}}

\copyrightyear{2022}
\acmYear{2022}
\setcopyright{acmcopyright}\acmConference[SBES 2022]{XXXVI Brazilian Symposium on Software Engineering}{October 5--7, 2022}{Virtual Event, Brazil}
\acmBooktitle{XXXVI Brazilian Symposium on Software Engineering (SBES 2022), October 5--7, 2022, Virtual Event, Brazil}
\acmPrice{15.00}
\acmDOI{10.1145/3555228.3555276}
\acmISBN{978-1-4503-9735-3/22/10}

\acmConference[SBES'2022]{XXXVI Simpósio Brasileiro de Engenharia de Software - Trilha de Ferramentas}{03--07 de Outubro de 2022}{Uberlândia, MG, Brasil}
\begin{document}

\title{GitHub Proxy Server: A tool for supporting massive data collection on GitHub}

\author{Hudson Silva Borges}
\affiliation{%
  \institution{Universidade Federal de Mato Grosso do Sul - UFMS}
  \country{Campo Grande - MS - Brasil}
}
\email{hudson.borges@ufms.br}

\author{Marco Tulio Valente}
\affiliation{%
  \institution{Universidade Federal de Minas Gerais - UFMG}
  \country{Belo Horizonte - MG - Brasil}
}
\email{mtov@dcc.ufmg.br}

\begin{translatedabstract}{english}
GitHub is the most popular social coding platform and widely used by developers and organizations to host their \textit{open-source} projects around the world. Besides that, the platform has a web API that allow developers collect information from public repositories hosted on it. However, collecting massive amount of data from GitHub can be very challenging due to existing restrictions and abuse detection mechanisms. In this work, we present a tool, called \textsc{GitHub Proxy Server}, which abstracts such complexities into a tool that is independent on operational system and programming language. We show that, using the proposed tool, it is possible to improve the performance of GitHub mining tasks without any additional complexities.

\vspace{1.5em}

\noindent\textbf{Ferramenta:} \url{https://github.com/gittrends-app/github-proxy-server}\\
\noindent\textbf{Apresentação:} \url{https://youtu.be/Dld9sK2lE1k}\\
\noindent\textbf{Licença:} MIT\\

\end{translatedabstract}

\begin{abstract}
GitHub é a plataforma de codificação social mais popular e amplamente utilizada por comunidades e empresas para hospedagem de projetos \textit{open-source}. Além disso, a plataforma conta com uma poderosa API que permite a pesquisadores coletarem informações públicas de projetos hospedados nela. Contudo, a coleta massiva de dados pode ser bastante desafiadora devido a limitações e mecanismos de detecção de abusos existentes. O presente trabalho apresentada uma ferramenta, chamada \textsc{GitHub Proxy Server}, que abstrai tais complexidades por meio de uma arquitetura independente de plataforma e linguagem de programação. Experimentos realizados com a ferramenta mostram que é possível melhorar o desempenho de tarefas de mineração do GitHub sem que complexidades adicionais sejam inseridas nos projetos.
\end{abstract}


\maketitle

\section{Introdução}

Nos últimos anos, GitHub tem sido utilizado como fonte primária de dados por pesquisadores e desenvolvedores interessados na melhoria do estado da arte e na construção de ferramentas de apoio ao desenvolvimento \textit{open-source}~\cite{cosentino2016findings, entinoIC17}. De fato, GitHub é a plataforma de codificação social mais popular e amplamente utilizada por desenvolvedores e empresas para hospedagem de seus projetos~\cite{borges2016understanding}. Alguns dos fatores que ajudam a explicar tal popularidade são as diversas funcionalidades disponíveis que vão muito além de simples ferramentas de controle de versão, como, por exemplo, acompanhar a atividade de outros desenvolvedores (\textit{follow}) e fóruns de discussão integrados (\textit{Discussion}) no próprio projeto~\cite{thung2013network, dabbish2012social, borges2016predicting, borges2018s}.

Além das funcionalidades sociais, o GitHub permite que seus usuários acessem seus serviços web por meio de APIs (\textit{Application Programming Interface}) públicas. Atualmente duas versão estão disponíveis: REST API (v3)\footnote{\url{https://docs.github.com/rest}} e GraphQL API (v4).\footnote{\url{https://docs.github.com/graphql}} Em ambas APIs é possível que usuários coletem informações públicas de quaiquers projetos hospedados na plataforma. Contudo, para acessar tais informações é necessário que seus usuários sigam regras e recomendações da própria plataforma, o que pode representar um grande desafio para aqueles que necessitam coletar uma quantidade massiva de dados~\cite{Gousios2017}. Por exemplo, cada usuário autenticado com um token tem o direito de realizar até 5.000 requisições por hora. Além disso, realizar requisições paralelas com o mesmo token pode bloquear o acesso do usuário temporariamente. De acordo com o GitHub, tais regras e recomendações são aplicadas para garantir a disponibilidade do serviço e um uso justo de seus serviços.\footnote{\url{https://docs.github.com/en/enterprise-cloud@latest/rest/overview/resources-in-the-rest-api\#rate-limiting}}

Com intuito de simplificar esse processo e tornar o processo de coleta mais transparente a desenvolvedores e pesquisadores que usam a API do GitHub, este trabalho apresenta uma ferramenta, chamada \textit{GitHub Proxy Server}, que abstrai todas as limitações e recomendações da plataforma GitHub por meio de uma arquitetura proxy independente de sistema operacional e linguagem de programação. A ferramenta tem como principais características: (i) suporte a múltiplos tokens de acesso; (ii) orquestramento automático de requisições simultâneas, (iii) balanceamento de carga, e (iv) configurações ajustáveis. Um estudo de caso com a ferramenta mostrou que a integração é simples com as principais bibliotecas disponíveis atualmente além de permitir que desenvolvedores utilizem os recursos computacionais disponíveis ao máximo.

Um estudo de caso com a ferramenta, mostra que a integração com bibliotecas existentes é bastante simples, bastando simplesmente modificar o endereço base de destino para a ferramenta proposta. Além disso, é possível otimizar o tempo gasto na coleta de dados ao utilizar processos paralelos sem que essa atividade viole as recomendações do próprio GitHub. 
 
A seguir é apresentada a organização deste artigo. Na Seção~\ref{sec:background} são detalhadas as limitações e restrições impostas pela plataforma GitHub em sua API, algumas alternativas à API do GitHub e como tais limitações impactam desenvolvedores e pesquisadores. Na Seção~\ref{sec:arquitetura} é detalhada a arquitetura proposta e na Seção~\ref{sec:ferramenta} a implementação da ferramenta. A Seção~\ref{sec:integracao} apresenta um estudo sobre como integrar a ferramenta proposta em ferramentas e atividades já existentes enquanto a Seção~\ref{sec:related} realiza uma comparação com ferramenta relacionadas. Por fim, a Seção~\ref{sec:conclusions} conclui o trabalho e apresenta direções futuras para a ferramenta.

\section{Coleta de dados no GitHub}
\label{sec:background}

Na literatura, a maioria dos estudos exploram os desafios envolvidos em pesquisas que envolvem mineração de dados do GitHub. Por exemplo, Kalliamvakou~\textit{et al.}~\cite{kalliamvakou2016depth} reportam as promesas e os perigos de se minerar repositórios GitHub com foco nas características dos projetos hospedados. Os autores também derivam um conjunto de recomendações a pesquisadores interessados em usar projetos hospedados na plataforma. 

Contudo, poucos estudos reportam os desafios técnicos envolvidos na coleta de dados do GitHub. Usuários que desejam coletar dados das APIs do GitHub (\textit{i.e.,} REST ou GraphQL) devem estar atentos quanto à limitação no número de requisições que cada usuário pode realizar. Usuários não autenticados com um token de acesso podem fazer um total de 60 requisições em uma hora enquanto usuários autenticados têm esse limite elevado a 5.000 requisições/hora/token. Ao ultrapassar esse limite usuários deverão aguardar o limite ser resetado para realizar novas requisições. Uma alternativa adotada por diversos usuários consiste em obter mais de um token de acesso, de diferentes usuários, para multiplicar a capacidade de coleta.

Além disso, o GitHub implementa uma série de mecanismos internos para detecção de abusos no uso de seus serviços. Usuários que não seguem as recomendações da plataforma podem ter seu acesso bloqueado temporariamente, como demonstra a seguinte mensagem da própria API: "\textit{You have triggered an abuse detection mechanism and have been temporarily blocked from content creation. Please retry your request again later}". Portanto, usuários devem estarem atentos às recomendações para não incorrer em problemas.

Por exemplo, não é recomendado que usuários realizem múltiplas requisições em paralelo para seus servidores. Por exemplo, não é recomendado que usuários enviem requisições simultâneas para dois endpoints diferentes (por exemplo, estrelas e issues) usando o mesmo token de acesso. Além disso, é recomendado que usuários adicionem um tempo mínimo entre as requisições. Atualmente a plataforma disponibiliza um guia de boas práticas com diversas dicas aos usuários de seus serviços.\footnote{\url{https://docs.github.com/en/rest/guides/best-practices-for-integrators}}

Uma alternativa adotada por pesquisados a fim de facilitar a tarefa de coleta de dados, é a adoção de datasets com dados previamente coletados. Gousios~\textit{et al.}~\cite{gousios2014lean} disponibiliza periodicamente dados coletados da API REST por meio do projeto \textsc{GHTorrent}. Já o projeto \textsc{GHArchive}\footnote{\url{https://www.gharchive.org/}} monitora, arquiva e compartilha todos eventos públicos lançados pelo GitHub~\cite{munaiah2017curating}. Contudo, o grande problema deste deste modelo de dados está relacionada à atualizações e completude dos dados. Por exemplo, até a escrita deste trabalho, os dados do \textsc{GHTorrent} foram atualizados 
em 06/2019 (três anos atrás).

\section{Arquitetura Proposta}
\label{sec:arquitetura}

\begin{figure}[ht]
    \centering
    \includegraphics[width=.75\columnwidth]{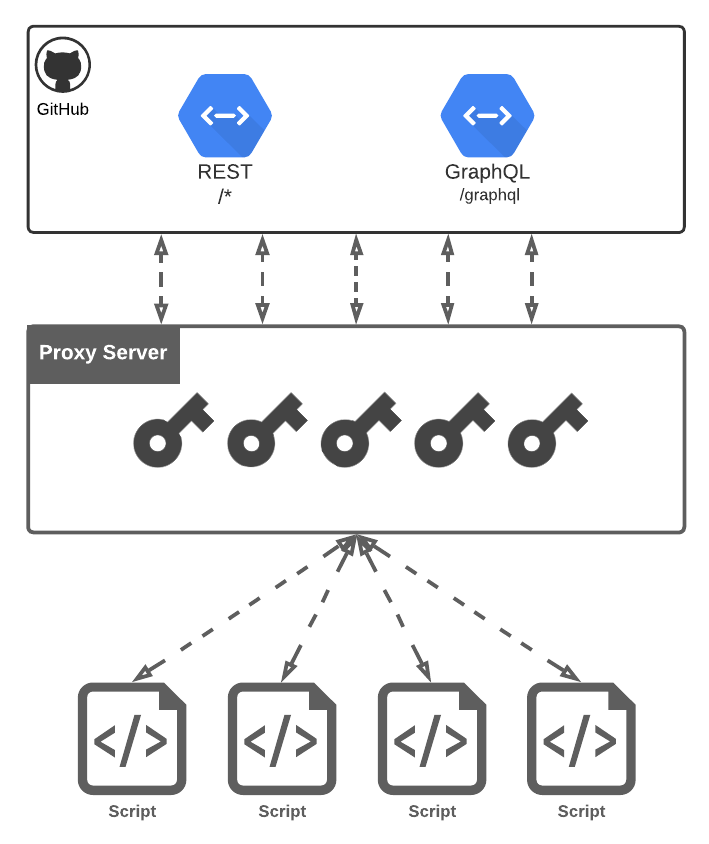}
    \caption{Arquitetura do \textsc{GitHub Proxy Server}}
    \label{fig:arquitetura}
\end{figure}

Com o propósito de ser uma ferramenta independente de plataforma e de linguagem de programação, \emph{GitHub Proxy Server} foi concebido para atuar como um \emph{proxy} entre os serviços do GitHub (i.e., REST API e GraphQL API) e aqueles usuários que necessitam fazer coletas massivas do dados. Essencialmente, na arquitetura proposta, o servidor \emph{proxy} é responsável por receber as requisições dos usuários e encaminhá-las aos serviços do GitHub de forma a evitar as restrições e minimizar as limitações originalmente impostas deixando seus usuários livres da responsabilidade de gerenciar os \emph{tokens} de acesso disponíveis. A Figura~\ref{fig:arquitetura} apresenta a arquitetura da ferramenta proposta. 

Para que o servidor \emph{proxy} execute corretamente os usuários devem prover, pelo menos, um token de acesso ao GitHub. Esse token pode ser gerado pelo próprio usuário na página do GitHub ou usar de tokens de acesso gerados por aplicações cadastradas no GitHub. Para cada um dos tokens fornecidos, o servidor cria uma \emph{worker} interno que é responsável por gerenciar e coordernar todas as requisições feitas a partir dele. Além disso, quando o servidor \textit{proxy} recebe uma requisição dos clientes, ele automaticamente encaminha essa requisição ao \emph{worker} que possui maior capacidade e disponibilidade de processá-la.

Na arquitetura proposta, cada requisição é recebida pelo servidor e o mesmo verificará se a requisição possui um token de acesso e um agente de usuário (\textit{a.k.a}, user-agent), informações que devem estar presentes em todas requisições aos serviços do GitHub. Caso algumas destas informações estejam ausentes, o \emph{worker} automaticamente preencherá com seu respectivo token e um valor padrão para o agente de usuário.

Assim, na arquitetura proposta, usuários poderão realizar as requisições necessárias sem maiores preocupações além de permitir que diversas aplicações clientes estejam em execução simultaneamente, fazendo assim um melhor aproveitamento dos recursos computacionais disponíveis. Por exemplo, um grupo de pesquisa ou desenvolvedores que compartilham de um conjunto de tokens de acesso e desejam coletar dados da plataforma podem criar uma instância única do servidor \textit{proxy} e usarem sem nenhum esforço adicional de configuração ou sincronização.

A seguir são apresentadas as principais funcionalidades da ferramenta proposta:

\vspace{1em}\noindent\textbf{Suporte a múltiplos tokens de acesso:} Os serviços do GitHub limitam o número de requisições que usuários podem realizar a seus serviços. Para que estes consigam ir além destes limites, é frequente que usem de vários tokens ao mesmo tempo. A ferramenta proposta permite que usuários forneçam diversos tokens e faz com que cada um deles atue como um cliente independente, fazendo a gestão do número de requisições disponíveis e tratando automaticamente eventuais problemas nas requisições. 

\vspace{.5em}\noindent\textbf{Orquestramento das requisições:} Ao realizar requisições simultaneas usando o mesmo token, o GitHub pode identificar e bloquear o acesso deste usuário. Para evitar tal situação, a ferramenta proposta cria uma fila de requisições para cada token, evitando que tal situação ocorra. Além disso, é possível adicionar um tempo de espera entre requisições para evitar chamadas sucessivas sem intervalo. Tal abordagem permite que o servidor evite o abuso no uso dos serviços do GitHub, assim como recomendado pela própria plataforma.

\vspace{.5em}\noindent\textbf{Balanceamento de carga:} Com objetivo de fazer o melhor aproveitamento possível das requisições disponíveis, é proposto uma função de balanceamento na distribuição das requisições recebidas para os \emph{workers} disponíveis. Assim, se houver \emph{workers} inativos, as requisições são encaminhadas diretamente. Senão, a requisição é encaminhada para o \emph{worker} com menor fila de requisições pendentes. Havendo mais de uma fila com o mesmo número de requisições pendentes, a requisição é encaminhada para a primeira com maior número de requisições disponíveis para o GitHub.

\vspace{.5em}\noindent\textbf{Configurações ajustáveis:} Todas as configurações adotadas pela ferramenta são personalizáveis com objetivo de atender melhor às necessidades dos usuários. Por exemplo, por padrão, o tempo entre requisições dos \emph{workers} é definido em 250 milissegundos. Contudo, usuários podem aumentar ou reduzir de acordo com suas necessidades. Além disso, é possível definir um limite máximo de uso para cada token, deixando um conjunto mínimo restante para os proprietários dos tokens de acesso usados.

\section{Ferramenta}
\label{sec:ferramenta}

\begin{figure}[ht]
    \centering
    \includegraphics[width=\columnwidth]{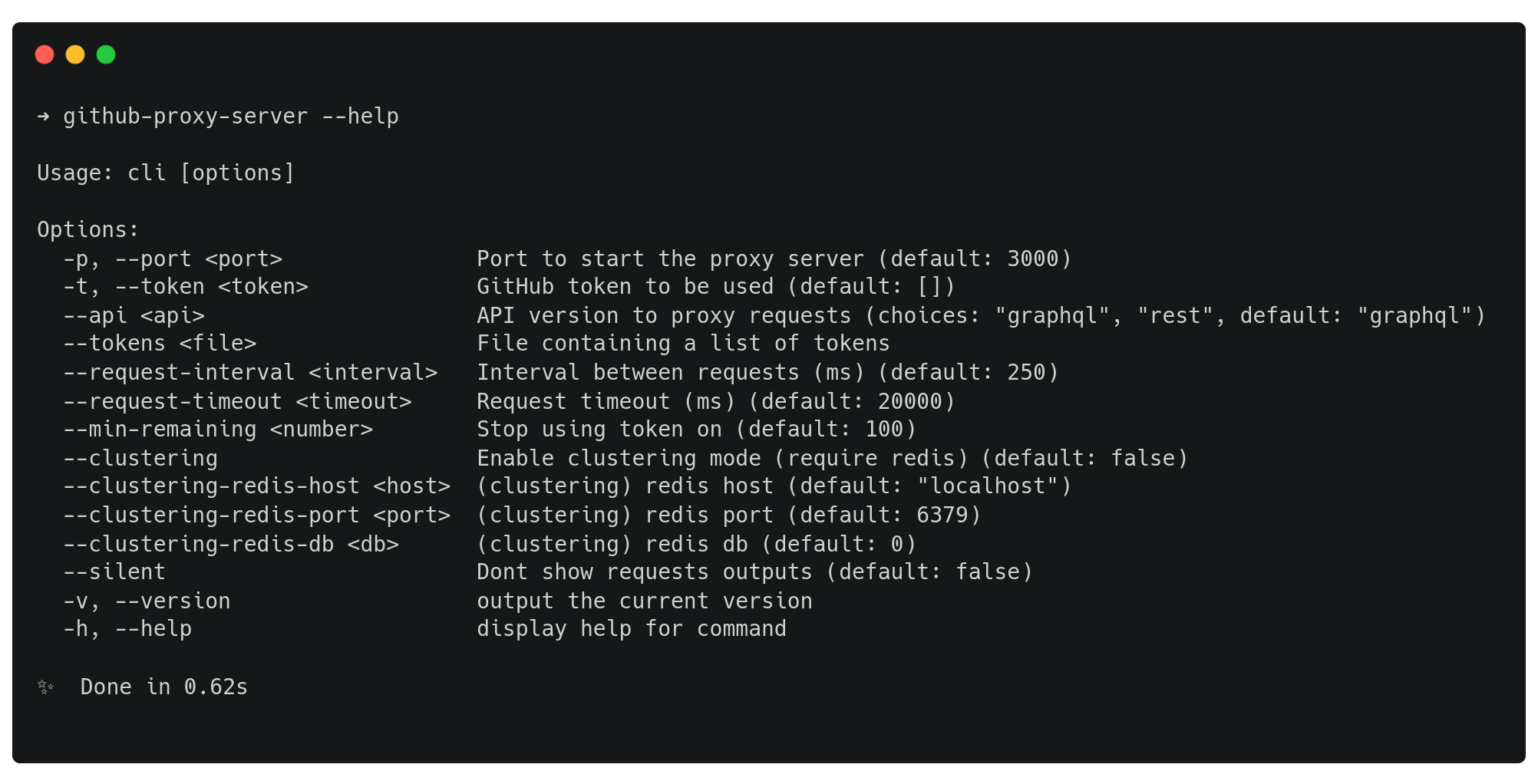}
    \caption{Interface da aplicação}
    \label{fig:help}
\end{figure}

Uma implementação da arquitetura proposta está disponível publicamente em \url{https://github.com/gittrends-app/github-proxy-server} sob a licença MIT.\footnote{\url{https://choosealicense.com/licenses/mit}} A ferramenta pode ser instalada e utilizada usando o gerenciador de pacotes NPM (Node Package Manager)\footnote{\url{https://www.npmjs.com/}} ou a partir dos executáveis disponíveis para cada uma das plataformas (Linux, Windows e MacOs). Atualmente, a ferramenta pode ser executada somente a partir do terminal e tem como requisito obrigatório que seus usuários forneçam os tokens de acesso do GitHub via argumentos, variáveis de ambiente ou arquivo de configuração. A Figura~\ref{fig:help} apresenta uma captura de tela com todas opções aceitas pela ferramenta.

Por padrão, a ferramenta está configurada para trabalhar com a API GraphQL, versão que sucede a API REST. Contudo, usuários podem selecionar qual versão ele deseja utilizar. Essa informação é necessária pois o controle das requisições é feita por serviço, ou seja, as contagens são independentes.

Também é possível configurar o intervalo entre as requisições (opção \textit{--request-interval}) que será adotado, o tempo máximo de execução de uma requisição (opção \textit{--request-timeout}) antes de ser automaticamente abortada, e o número mínimo de requisições que a aplicação deve manter sem utilizá-las (opção \textit{--min-remaining}). Por fim, os usuários podem optar pela construção de clusters de servidores (opção \textit{--clustering}) para expandir a capacidade dos mesmos mantendo a sincronia entre eles.

\begin{figure}[ht]
    \centering
    \includegraphics[width=0.85\columnwidth]{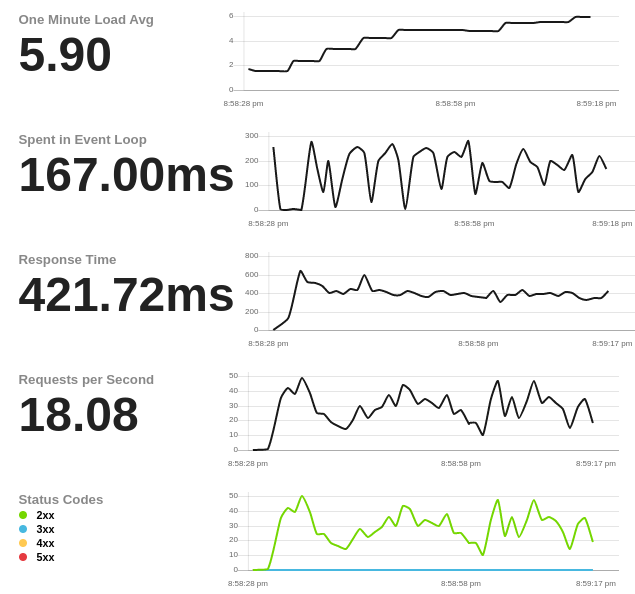}
    \caption{Monitoramento de atividades}
    \label{fig:monitoring}
\end{figure}

A ferramenta também possui um serviço de \textit{logging} e um monitor de atividades que permite seus usuários monitorem as requisições que passam por ele, incluindo um resumo dos resultados (i.e., \textit{status code}) das requisições aos serviços do GitHub (Figura~\ref{fig:monitoring}). O monitoramento do uso permite aos seus usuários ajustarem de forma mais adequada os parâmetros da ferramenta.

\section{Integração com outras ferramentas}
\label{sec:integracao}

Como descrito anteriormente, a ferramenta proposta foi concebida para ser independente de plataforma e linguagem de programação. Além do mais, ela tem por objetivo permitir que qualquer usuário possa utilizar da ferramenta sem a necessidade de grandes modificações em seus aplicações. 

Atualmente diversas bibliotecas estão disponíveis para desenvolvedores interessados em acessar os serviços do GitHub por meio de suas linguagens de programação preferidas. Inclusive, o próprio GitHub provê uma lista de bibliotecas disponíveis em sua documentação oficial.\footnote{\url{https://docs.github.com/pt/rest/overview/libraries}} Neste trabalho, foram analisadas inicialmente as três bibliotecas oficiais da própria equipe do GitHub: (i) octokit.rb\footnote{\url{https://github.com/octokit/octokit.rb}}, para Ruby, (ii) octokit.net\footnote{\url{https://github.com/octokit/octokit.net}}, para .NET, e (iii) octokit.js\footnote{\url{https://github.com/octokit/octokit.js}}, para JavaScript. Nestas três bibliotecas foram observadas formas de definir a URL de destino das requisições, possibilitando assim a integração das aplicações com a ferramenta proposta.

\begin{figure}[ht]
    \centering
    \includegraphics[width=0.9\columnwidth]{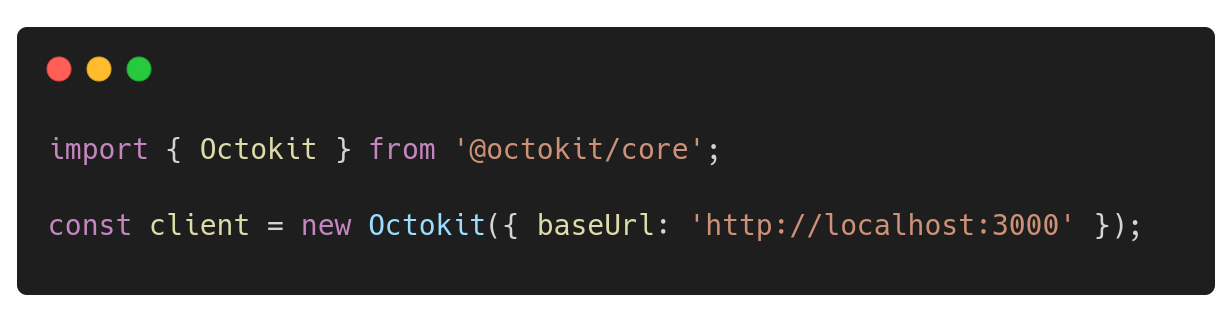}
    \caption{Integração com \textsc{octokit.js}}
    \label{fig:integração}
\end{figure}

Por exemplo, a Figura~\ref{fig:integração} mostra como um desenvolvedor pode configurar a biblioteca \textsc{octokit.js} para usar o \textsc{GitHub Proxy Server}. Basicamente, usuários devem direcionar as requisições da biblioteca para o servidor proxy modificando a propriedade \textit{baseUrl} no construtor do cliente.

Considerando bibliotecas de outras linguagens, também foi observado que diversas delas possuem suporte ao uso de diferentes URLs destino ou configuração de proxy interno. Por exemplo, a biblioteca \textsc{PyGitHub}\footnote{\url{https://github.com/PyGithub/PyGithub}}, a mais popular para a linguagem Python com mais de 5k estrelas, permite que usuários usem diferentes URLs como base para as requisições a partir do parâmetro \textit{base\_url}.

\begin{figure}[ht]
    \centering
    \includegraphics[width=0.9\columnwidth]{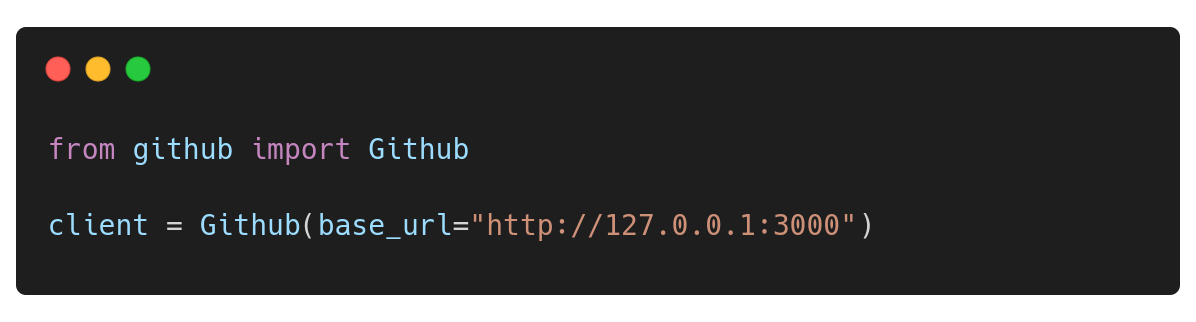}
    \caption{Integração com \textsc{PyGithub}}
    \label{fig:integração2}
\end{figure}

\subsection{Exemplo de Uso}
\label{sec:usage}

Para ilustrar e medir o impacto da ferramenta em atividades de mineração, propõe-se a seguinte atividade: \textit{Realizar a busca pelos repositórios mais populares do GitHub, em número de estrelas, coletar suas issues, releases, tags e stargazers usando a API REST e armazenando os dados obtidos em uma base de dados não relacional}. 

Neste trabalho, será comparado o desempenho obtido nessa atividade usando: (i) a ferramenta proposta e (ii) realizando as requisições diretamente ao serviço do GitHub (\textit{i.e.}, sem a ferramenta). Também será avaliado o desempenho ao utilizar um único token de acesso e utilizando múltiplos tokens.

\begin{figure}[ht]
    \centering
    \includegraphics[width=0.9\columnwidth]{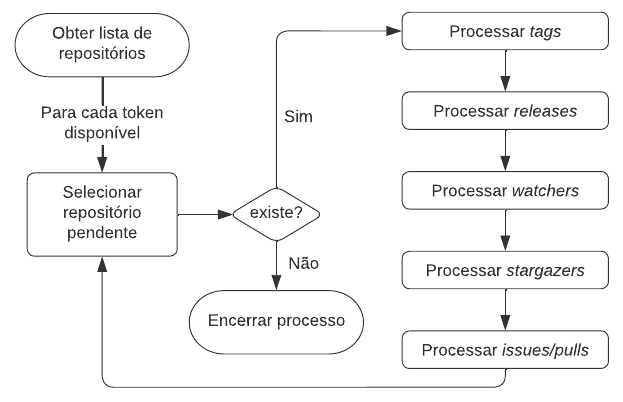}
    \caption{Processamento sem a ferramenta}
    \label{fig:fluxograma1}
\end{figure}

Considerando que, ao realizar as requisições diretamente aos serviços do GitHub, os desenvolvedores estão sujeitos a todas limitações e restrições da plataforma, neste exemplo de uso optou-se por realizar as requisições sequencialmente, com um intervalo após as requisições de 50 milissegundos, para cada token disponível. Como consequência a coleta dos recursos também foi sequencial. A Figura~\ref{fig:fluxograma1} ilustra o processo adotado para a atividade de coleta sem usar da ferramenta proposta. No modelo de processamento proposto, é garantido que somente uma requisição seja feita a cada momento por cada token disponível.

\begin{figure}[ht]
    \centering
    \includegraphics[width=\columnwidth]{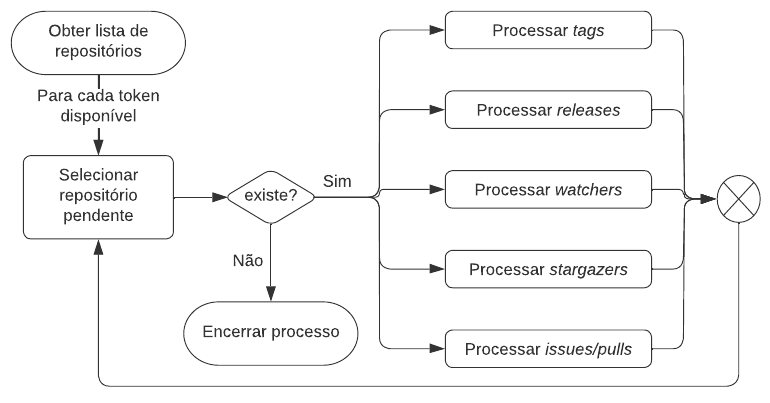}
    \caption{Processamento usando o \textsc{GitHub Proxy Server}}
    \label{fig:fluxograma2}
\end{figure}

Já para o processo de coleta de dados usando a ferramenta proposta (Figura~\ref{fig:fluxograma2}), os scripts foram modificados para realizarem processamentos paralelos. Neste caso, requisições paralelas não implicam em eventuais problemas porque a ferramenta proposta, ao receber tais requisições, é capaz de processá-las sem inferir as restrições de uso dos serviços do GitHub. Neste modelo, foram utilizados três processos paralelos para cada token disponível. Além disso, cada processo pode ter até seis requisições simultâneas a cada instante (uma para cada recurso do GitHub em processamento).

Os scripts utilizados neste exemplo de uso estão disponíveis na branch \textsc{cbsoft-tools-2022} sob o diretório \textit{benchmark} do repositório do projeto. Além disso, os dados da análise reportada a seguir estão presentes no mesmo diretório assim como o arquivo \textsc{R Markdown}\footnote{\url{https://bookdown.org/yihui/rmarkdown/}} para reprodução dos resultados.

Para a execução dos testes, foi alugada uma máquina virtual em nuvem na Amazon Web Services (AWS) com 4GB de memória RAM, 2 vCPUs e um disco sólido de 80GB.\footnote{\url{https://aws.amazon.com/pt/lightsail/}} O ambiente foi configurado com a última versão disponível do Node.js\footnote{\url{https://nodejs.org/}} e o banco de dados MongoDB\footnote{\url{https://www.mongodb.com/}}. Todos os testes foram executados individualmente sob as mesmas condições de execução.

\subsubsection{Performance com único token de acesso}
\begin{figure}[ht]
    \centering
    \includegraphics[width=\columnwidth,trim={0 0 0 2cm},clip]{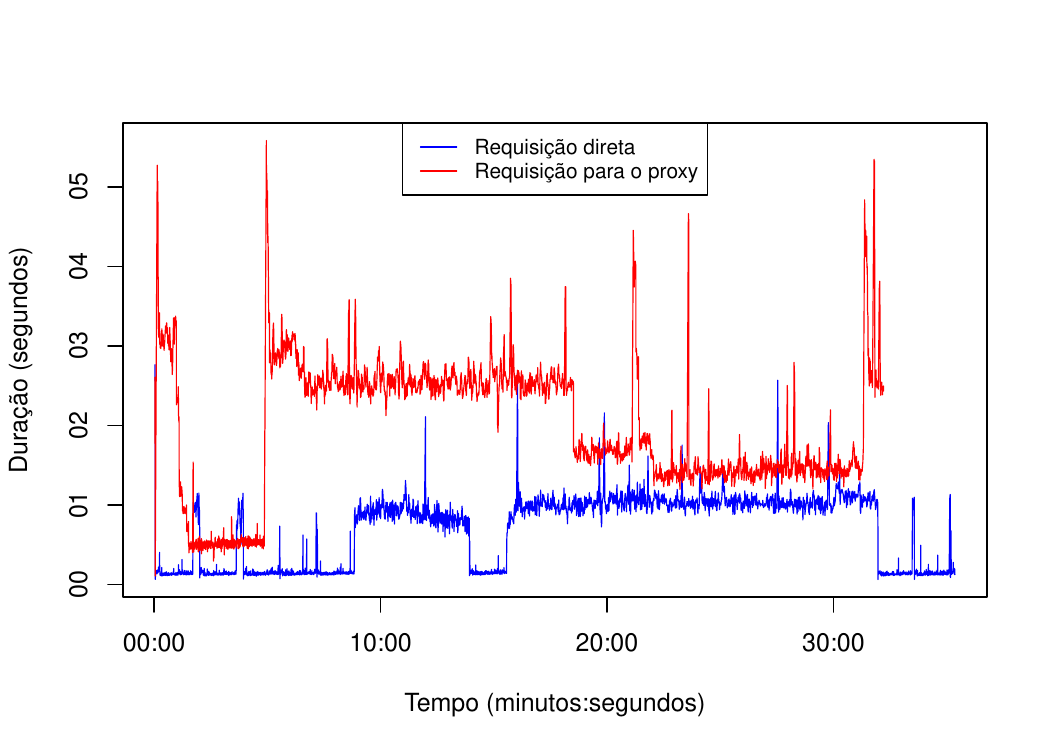}
    \caption{Linha do tempo das requisições}
    \label{fig:exemplo1}
\end{figure}

A Figura~\ref{fig:exemplo1} mostra o tempo gasto em cada requisição realizada para ambas as fontes. Como as requisições diretas aos serviços do GitHub foram feitas de forma sequencial, é possível observar que a duração das requisições foi bem pequena. Além disso, observou-se também que alguns endpoints da API (e.g., issues) demandam mais tempo para responder do que outros (e.g., stargazers). O tempo total para que todas requisições fossem realizadas foi de 35 minutos e 22 segundos.

Já as requisições realizadas através do servidor proxy apresentaram uma duração mais longa. Isso é justificado pelo fato da ferramenta proposta enfileirar requisições paralelas. Contudo, observou-se também que o mesmo número de requisições foi feito em 32 minutos e 13 segundos. Um dos motivos observados que ajudam a justificar os resultados obtidos, é melhor aproveitamento dos recursos computacionais disponíveis ao realizar processamentos paralelos. Por exemplo, tarefas de escrita dos dados no disco e requisições de rede podem ser facilmente intercaladas pelos sistemas operacionais.

\subsubsection{Performance com múltiplos tokens de acesso}
\begin{figure}[ht]
    \centering
    \includegraphics[width=\columnwidth,trim={0 0 0 2cm},clip]{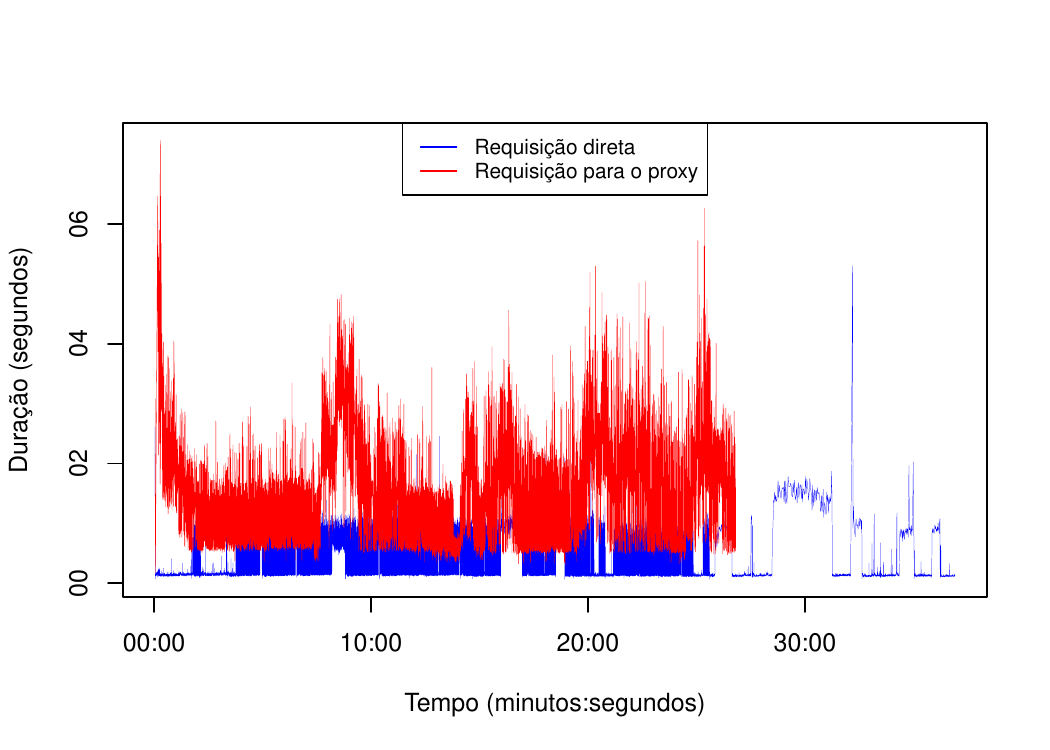}
    \caption{Linha do tempo das requisições (três tokens)}
    \label{fig:exemplo2}
\end{figure}

A Figura~\ref{fig:exemplo2} mostra a linha do tempo ao realizar a tarefa proposta usando três tokens de acesso. Em ambos os casos, é possível observar que existe uma concentração maior de requisições ao longo do tempo. Mas, mais uma vez, o uso do servidor proxy para intermediação das requisições permitiu que a coleta fosse realizada em um tempo consideravelmente menor. Por meio de requisições diretas, o tempo necessário para processar todas requisições foi de 36 minutos e 55 segundos, valor bem próximo ao obtido anteriormente. Usando o servidor proxy o tempo foi conideravelmente menor, de 26 minutos e 49 segundos. Portanto, desenvolvedores podem beneficiar-se bem mais ao utilziar a ferramenta com múltiplos tokens de acesso.

\section{Ferramentas relacionadas}
\label{sec:related}

Na literatura, diversas ferramentas para análise de repositórios \textit{git} foram propostas. Por exemplo, a ferramenta \textsc{PyDriller} permite desenvolvedores clonarem e extrairem informações sobre \textit{commits}, desenvolvedores, e outros metadados diretamente dos projetos~\cite{Spadini2018}.
Contudo, até o momento da escrita deste artigo, nenhuma outra ferramenta \textit{open-source} de mesmo propósito à apresentada neste trabalho foi encontrada. A ferramenta proposta tem por objetivo auxiliar pesquisadores a conduzirem coletas massivas de dados independente da plataforma ou linguagem de programação adotada pelos mesmos diretamente do GitHub.

Contudo, bibliotecas para certas linguagens de programação podem oferecer funcionalidades semelhantes às propostas nesta ferramenta. Por exemplo, a biblioteca oficial do GitHub para a linguagem JavaScript, \textsc{octokit.js}, fornece um conjunto de funcionalidades para interagir com as APIs REST e GraphqQL por meio de plugins. Especificamente, o plugin \textsc{octokit/plugin-throttling.js}\footnote{\url{https://github.com/octokit/plugin-throttling.js}} implementa todas as recomendações de boas práticas do próprio GitHub com objetivo de previnir os mecanismos de detecção de abusos. Conudo, a biblioteca não permite adicionar vários tokens de acesso como a ferramenta proposta.

A fim de verificar se outras bibliotecas também oferecem tais recursos, foi conduzida uma breve análise sobre a documentação das bibliotecas de terceiros para JavaScript listadas na documentação da própria plataforma.\footnote{\url{https://docs.github.com/pt/rest/overview/libraries}} Observou-se que nenhuma das quatro bibliotecas implementa funcionalidades equivalentes. O mesmo comportamento foi observado em bibliotecas de outras linguagens de programação. Portanto, isso ressalta a importância da ferramenta proposta.

\section{Considerações finais}
\label{sec:conclusions}

O GitHub tem sido utilizado como fonte primária de dados por diversos pesquisadores na condução de suas pesquisas. Dentre os fatores que ajudam explicar tal fenômeno estão a popularidade da plataforma na comunidade \textit{open-source} e também uma poderosa API que permite que desenvolvedores e pesquisadores possam acessar informações nela disponíveis. Contudo, o acesso a tais informações não é trivial e seus usuários estão sujeitos a restrições que podem inviabilizar a coleta massiva de dados.

Neste artigo é apresentada uma ferramenta, chamada \textsc{GitHub Proxy Server}, que tem como objetivo abstrair as restrições e recomendações de acesso aos serviços do GitHub de forma a tornar o processo totalmente transparente para desenvolvedores e pesquisadores que desejam coletar dados da plataforma. Avaliações conduzidas com a ferramenta mostraram que a integração com bibliotecas e outras ferramentas existentes é bastante simples. Além disso, ao permitir o uso de concorrência de requisições, os usuários da ferramenta podem otimizar suas atividades para melhor aproveitamento dos recursos computacionais disponíveis. Por fim, considerando que a autenticação é feita por meio da modificação do \textit{header} das requisições, a solução proposta tem o potencial de suportar, inclusive, futuras versões da API do GitHub.

Como trabalhos futuros, planeja-se adicionar uma funcionalidade de obtenção automática de tokens de acesso. Essa obtenção visa facilitar a configuração da ferramenta e facilitar a tarefa que atualmente deve ser realizada manualmente pelos usuários. Pretende-se também avaliar uma estratégia de ajuste automático de parâmetros com base no uso. Atualmente tais valores são configurados manualmente ao iniciar a aplicação.


\bibliographystyle{ACM-Reference-Format}
\bibliography{references}

\end{document}